\documentclass{appolb}
\usepackage{epsfig}
\usepackage[T1]{fontenc}
\begin{document}
\pagestyle{plain}
\title{$C_5^A$ form factor from ANL experiment%
%single author
%\thanks{Presented at the 45th Winter School in Theoretical Physics ``Neutrino Interactions: from Theory to Monte %Carlo Simulations'', L\k{a}dek-Zdr\'oj, Poland, February 2--11, 2009.}%
%multiple authors:
\thanks{Presented by Krzysztof Graczyk at the 45th
Winter School in Theoretical Physics {\it Neutrino Interactions: from Theory to Monte Carlo Simulations}, L\k{a}dek-Zdr\'oj, Poland, February 2--11, 2009.}%
}
\author{Krzysztof M. Graczyk$^\dag$,
D. Kie\l czewska$^\ddag$, and Jan T. Sobczyk$^\dag$
%\address{and their affiliation}
\address{$^\dag$Institute of Theoretical Physics,  University of Wroc\l aw, \\ pl. M. Borna 9, 50-204, Wroc\l aw, Poland}
\address{$^\ddag$ A. Soltan Institute for Nuclear Studies \\ Hoza 69, 00-681 Warsaw, Poland }
}
\maketitle
\begin{abstract}
$C_5^A(Q^2)$ axial form factor is extracted from the ANL neutrino-deuteron
scattering data with deuteron structure effects taken into consideration. The best fit of
the $C_5^A(Q^2)$ axial form factor is obtained assuming dipole parametrization with $C_5^A(0)=1.13\pm0.15$ and
$M_A=0.94\pm 0.08$ GeV.
\end{abstract}

\section{Introduction}

A knowledge of the cross sections for single pion production in neutrino-nucleon scattering is an
important ingredient of the long base-line oscillation experiment analyses (see experiments: K2K \cite{k2k} and T2K \cite{t2k}).  In particular,
prediction of the cross sections for $1\pi^0$ production in neutral current neutrino-nucleon scattering has great
importance for the estimation of the background for observation of the $\nu_\mu\to\nu_e$ neutrino oscillation. The reason is that the $\pi^0$ decay signal into two photons can be mislead
in the detector with the electron shower.

More than twenty years ago two bubble chamber experiments: 12ft detector at Argonne National
Laboratory (ANL) \cite{Barish,Radecky}   and 7ft detector at Brookhaven National Laboratory (BNL) \cite{Kitagaki}
had collected enough data to investigate the $W$, $Q^2$ and energy dependence of the differential
and total cross sections for neutrino-deuteron scattering. The ANL and BNL experimental measurements can still serve as an important source of information about cross sections for
quasi-elastic and single pion production in neutrino scattering.
The nuclear effects (deuteron structure corrections) are relatively
easy to consider, therefore analysis of these data allows to study neutrino-nucleon reactions.

In this paper a part of the re-analysis \cite{nasza_praca} of the neutrino-deuteron
scattering data is presented. We re-examine the single pion production in neutrino deuteron
interactions. Our aim is re-extraction of the $C_5^A(Q^2)$ axial form factor from the $1\pi$ data.
In this article we discuss only the data collected at Argonne National Laboratory. In the longer contribution
simultaneous analysis of both ANL and BNL data is described.

The presentation is organized as follows in Sect. \ref{section_Single_pion_production} the model for
$\Delta(1232)$ excitation is shortly introduced. Sect. \ref{section_reanalysis_anl} contains description of the ANL data and details of the
statistical analysis. Sect. \ref{section_result} includes our final results and conclusions.

\section{Single pion production \label{section_Single_pion_production}}

We consider the following process:
\begin{equation}
\label{rection}
\nu(k) + p(p) \to \mu^-(k')  + \Delta^{++}(p') \{ \rightarrow \pi^+(l) + p(r)\}.
\end{equation}
By the $k$ and $k'$ the neutrino and muon four-momenta are denoted. The incoming nucleon
momentum is denoted by $p$, while $l$ and $r$ are the final pion and proton momenta. The four momentum transfer
is defined as follows $Q^2= -(k-k')^2$. The total hadronic momentum is $p'=l+r$.

One of the simplest way to describe above reaction is to apply the Adler-Rarita-Schwinger
formalism \cite{Schreiner:1973mj}. In this description the scattering amplitude is given by the contraction of the lepton $j_{lep}^\mu$
with the hadronic
$\left<\Delta^{++},p'\right|{\mathcal{J}_\mu^{CC}}\left| p\right>$ currents. The hadronic current has Vector-Axial structure,
where the vector contribution is modeled by three unknown vector form factors $C_3^V$, $C_4^V$ and $C_5^V$. For these
form factors we apply recent fits from Ref. \cite{Lalakulich:2006sw}.

In general the axial contribution to the hadronic
current is given by four form factors, however, additional constrains, motivated by Adler model \cite{Adler} and
PCAC hypothesis, reduce the number of unknown functions to the one: $C_5^A(Q^2)$.

In the present analysis we consider the dipole parametrization of the $C_5^A(Q^2)$ axial form factor:
\begin{equation}
C_5^A(Q^2) = C_5^A(0) \left(1  +\frac{Q^2}{M_A^2} \right)^{-2}.
\end{equation}
The axial mass $M_A$ is a parameter to fit. We distinguish  two fit cases:
\begin{itemize}
\item[(i)]  the value of $C_5^A(0)$ is established by PCAC and equals $1.15$ \cite{BarquillaCano:2007yk};
\item[(ii)] $C_5^A(0)$ is treated as a free parameter.
\end{itemize}

\section{Re-analysis of the ANL data \label{section_reanalysis_anl}}

A subject of our analysis is the differential cross section data
\begin{equation}
\frac{d\sigma  }{ d Q^2} (Q^2_i) \equiv \sigma_{exp}(Q^2_i)
\end{equation}
which was published in Ref. \cite{Radecky}. In the reaction (\ref{rection}) the dominant
contribution  comes from the excitation of the $\Delta(1232)$ resonance. The nonresonant background can be neglected.

The proper analysis of the (\ref{rection})  process requires taking into consideration the deuteron structure effects.
In this work we follow the approach proposed in Ref. \cite{AlvarezRuso:1998hi}. From the practical point of view,
in order to obtain the differential cross section for neutrino-deuteron scattering,
we multiply neutrino-\textit{free nucleon } $\sigma(Q^2_i)$ formula by
$$R(Q^2)= \sigma(\nu d \to \mu^- \pi^+ p + n )/\sigma(\nu p \to \mu^- \pi^+ p)$$
correcting function. It is obtained from Ref. \cite{AlvarezRuso:1998hi}
(for more details see. Ref. \cite{nasza_praca}).

To analyze the data we apply the standard $\chi^2$ approach. Besides statistical and non-correlated systematical
uncertainties in each $Q^2_i$ bin, we take into account the overall uncertainty of the neutrino beam. It introduces to the standard $\chi^2$ formula additional normalization term:
\begin{equation}
\label{chi2_sigma} \chi^2=\sum_{i=1}^{n_{}}
\left(\frac{\sigma_{th}(Q^2_i) - p
\sigma_{ex} (Q^2_i)}{ p \Delta \sigma_i} \right)^2 +
\left(\frac{p -1}{ r }\right)^2,
\end{equation}
where $\sigma_{th}(Q^2_i)$ and  $\sigma_{ex}$ are the theoretical
and experimental values of the differential cross sections.  $\Delta \sigma_i$ denotes error in each $Q^2_i$ bin.
$r$ is the uncertainty of the neutrino flux and $p$ is the normalization
parameter which is going to be fitted.

In order to compute the theoretical values of the cross section the following kinematical cuts are imposed:
for the neutrino energy $E\in (0.5,6)$~GeV, and for the hadronic
invariant mass $W<1.4$~GeV. The data covers the range in $Q^2$ from $Q^2=0.01$~GeV$^2$ to $Q^2=1$~GeV$^2$.

The theoretical formula for the differential cross section in a given $Q^2$ bin is
the following:
\begin{eqnarray}
\sigma_{th}(Q^2_i) &=&   \int_{Q^2_{i,lo}}^{Q^2_{i,up}}  \frac{d Q^2}{\Delta Q^2_i}
\int_{E_{lo}}^{E_{up}} \frac{dE}{\Psi} \Phi(E) \int_{M+m_\pi}^{1.4~GeV}dW \sigma_{th}(E,Q^2,W)
\end{eqnarray}
with $Q^{2}_{i,lo} = Q^2_i - \Delta Q^2_i/2$, $Q^{2}_{i,up} = Q^2_i + \Delta Q^2_i/2$.
\begin{equation}
\Psi = \int_{E_{lo}}^{E_{up}} dE \Phi_{ANL}(E),
\end{equation}
where $E_{lo}=0.5$~GeV and $E_{up}=6$~GeV.

As it was discussed in Ref. \cite{Radecky}  the normalization
uncertainty of the total cross section, due to the lack of knowledge about the flux
is estimated to be 15\% for $E\in (0.5, 1.5)$~GeV  and 25\% for
$E>1.5$~GeV. Therefore in our discussion, we assume the average
overall normalization uncertainty to be 20\%, $r = 0.20$.

\begin{table}
\centering{ \begin{tabular}{|c|c|c|c|c|}
\hline
 &   &  & &   \\
 & $M_A$ [GeV] & $C_5^A(0)$ & $p$ & $\chi^2/\mathrm{NDF}$  \\
 &   &  & &  \\
\hline
 &   &  & &   \\
free target & 0.948 $\pm$ 0.074     &             &   1.15 $\pm$ 0.09   & $1.50/7$  \\
 &   &  & &   \\
\hline
 &   &  & &   \\
free target & 0.939 $\pm$ 0.082     &   1.04 $\pm$ 0.14   &   1.02 $\pm$ 0.19   & $0.94/6$  \\
 &   &  & &  \\
\hline
 &   &  & &   \\
deuteron & 0.937 $\pm$ 0.075     &      &   1.03 $\pm$ 0.09   & $0.81/7$  \\
 &   &  & &   \\
\hline
 &   &  & &   \\
deuteron & 0.936 $\pm$ 0.077     &  1.13$\pm$0.15    &   1.02 $\pm$ 0.19   & $0.80/6$  \\
 &   &  & &   \\
\hline
  \end{tabular}}
\caption{The obtained numbers for $C_5^A(Q^2)$ and $M_A$ resulting from the fitting procedure. \label{table_dipole_results}
}
\end{table}
\begin{figure}\centering{
\includegraphics[width=\textwidth, height=7cm]{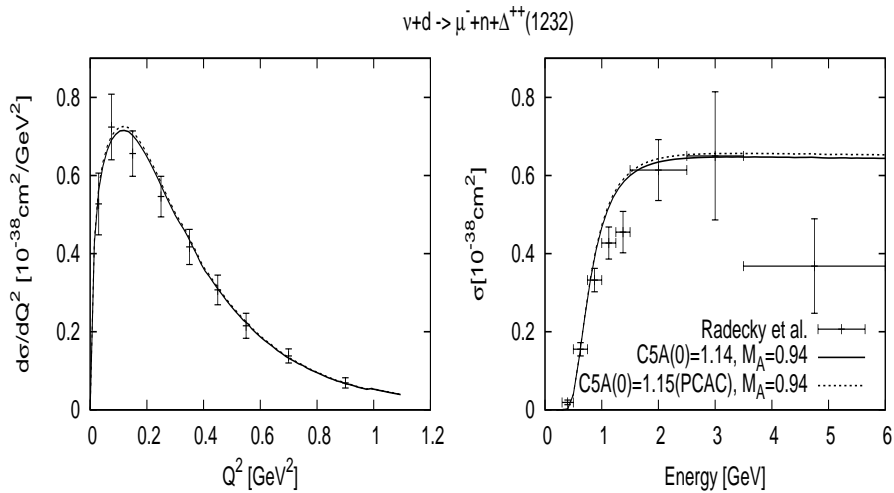}}
\caption{ The differential (left panel) and total (right panel) cross sections for
$\nu+d \to \mu^-+p+n + \Delta^{++}(1232)$ reaction measured at ANL experiment. The experimental points are
taken from \cite{Radecky}. The solid line denotes cross section computed with $C_5^A(0)=1.13$ and $M_A=0.936$ GeV.
The dashed line represents cross section computed with $C_5^A(0)=1.15$(fixed) and $M_A=0.937$ GeV.
Both fits were obtained with accounting deuteron structure effects. The cut $W<1.4$~GeV on hadronic invariant mass
is imposed. The normalization parameter is $p=1$. In the left panel in order to compute cross section the deuteron structure correction is imposed.
\label{fig_anl_dif}}
\end{figure}

\section{Results and conclusions \label{section_result}}

In Tab. \ref{table_dipole_results} the results of our fitting procedure are shown. In the analysis we consider four different cases.
We start the fitting procedure without accounting deuteron correction. The results are shown in the first two rows
of the Tab. \ref{table_dipole_results}).
Then the deuteron corrections are taken into consideration (look at the third and fourth rows of the Tab.
\ref{table_dipole_results}).

The comparison of our fits with the ANL data is shown in Fig. \ref{fig_anl_dif}, where both the differential and
total cross sections are plotted.

Our final result (with deuteron effects) is :
\begin{equation}
M_A=0.936\pm 0.077\;\;\mathrm{GeV}, \quad C_5^A(0) = 1.13\pm0.15.
\end{equation}

As it can be noticed for all fits the obtained values of $\chi^2/NDF$ (NDF -- number of degrees of freedom)
are very good. We observe that the axial mass,
which is responsible for the shape of $d\sigma /dQ^2$, is not affected by the
deuteron corrections. The deuteron nuclear effects affect mainly the normalization of the cross section,
therefore they are compensated by the normalization parameter $p$.

The largest suppression of the $\sigma(Q^2)$, due to deuteron effects appears in low-$Q^2$ (below 0.1~GeV$^2$). That
is why the fit of $C_5^A(0) $ without deuteron correction is relatively low ($\approx1.04$). When the deuteron effects are
accounted the obtained value is $\approx1.13$ which is very close to the PCAC value. However, both results for $C_5^A(0)$
are always in agreement with the PCAC value.

The fits obtained for the ANL data turns out to be quite similar to those obtained from the simultaneous fit to both the ANL and BNL data. The full
presentation of this analysis will be published in Ref. \cite{nasza_praca}.

\section*{Acknowledgements}
The authors were supported by the grant:  35/N-T2K/2007/0 (the project number DWM/57/T2K/2007).

\end{document}